# Non-local boundary conditions and internal gravity wave generation

V.V.Bulatov, Y.V.Vladimirov
Institute for Problems in Mechanics
Russian Academy of Sciences
Pr.Vernadskogo 101-1, 119526 Moscow, Russia
bulatov@index-xx.ru

**Abstract.**

This work focuses on the mathematical modeling of wave dynamics in a stratified medium. Non-local absorbing boundary conditions are considered based on the two following assumptions: (i) a linear theory can be applied at large distances from perturbation sources; and (ii) there are no other sources of wave disturbance outside the mixing zone in the stratified medium. The boundary conditions considered in this paper allowed us to describe the diverging internal gravity waves generated by the mixing region in a stratified medium.

**Introduction.**

As a rule in perturbation nonlocal source immediate neighborhood, for example near turbulent wakes arising from said sources movement, strong mixing zones, vertical formations hydrodynamics equations are essentially nonlinear and it is necessary to recourse to tedious numerical calculations to describe a near field. On the other hand for calculation of internal gravity waves long-distance propagation direct numerical calculations (for example by the finite-difference method) are unfeasible. However in the far zone interesting fields are relatively small-amplitude and usually it is possible to describe them with linear equations. Besides it is considered to be universally accepted that effects of viscosity, medium rotation and its compressibility are negligible and do not affect on internal gravity far-distance propagation. Therefore the far zone wave field as it is shown in [1-5], is described by relatively simple analytical formulae. Herewith initial or boundary conditions should be defined from either far field numerical calculation results with account of hydrodynamics nonlinear equations or from extremely evaluating (semiempirical) considering making it possible to approximate near field with some system of perturbation moving sources [1-5].

We shall assume, that perturbation source moves at a constant speed $V$ along the axis $x$ in negative direction ($x, y$ are horizontal coordinates, $z$ is vertical one). With neglect of nonstationary processes related for example to vortex periodic excitation while a source streaming in transient period, related to perturbation source motion start, the required field will be the function of variables $\xi = x + Vt$, $y$, $z$. It is possible to set the boundary problem for far field for example in the following manner:

The problem A. The planes $y = \pm y_0$ are taken, the perturbation source moves in the "passage" formed by these plane. With $|y| < y_0$ the field is numerically calculated, from this solution with $y > y_0$ (for definiteness) velocity $(U_1, U_2, W)$ and elevation $\xi$, component values are determined as functions of $\xi = x + Vt$ and $z$. Of these four functions only two appear to be independent ones (for example elevation $\varsigma$ and plane $y = y_0$ normal velocity $U_2$ component). No-source in the region $y > y_0$ condition imposes nonlocal constraint between these functions, it is sufficient to only set elevation $\varsigma$ (or velocity $U_2$ component), and one boundary condition appears to be sufficient to find the field in the region $y > y_0$.

The problem B. The plane $\xi = x + Vt = \xi_0$ following the perturbation source is taken, where $\xi_0$ is so large, that all effects of nonlinearity, viscosity and etc. are negligible with $\xi > .\xi_0$, i.e. with $\xi > .\xi_0$ the required field would be described by linear equations and would have no sources. The wave field with $\xi < \xi_0$ is calculated by numerical methods, the field with $\xi > \xi_0$ is calculated in linear approximation of boundary values on the plane $\xi = \xi_0$. No-source condition with $\xi > \xi_0$ imposes well defined nonlocal constraint on four functions describing the field (velocity three components and elevation), the required field in the region $\xi > \xi_0$ is constructed by three set independent functions (for example by elevation and velocity components $(U_2, W)$ in the plane $\xi = \xi_0$).

With far field definition problem converting to problems A and B near field calculation is the necessary stage – up to now unsolved in corpore problem, although it is possible to expect, that numerical methods development will make it possible in the foreseeable future to obtain its solution. It is possible to consider the third approach, which make it possible to evaluate internal gravity waves amplitude and phase structure.

The problem C. The near field is approximated by some source system, where the far field is defined as the field excited by these sources. The internal gravity waves arising with the perturbation source streaming are approximated as the field excited by horizontally oriented mass dipole or sources and drains combination, where distribution of these sources is taken from the problem of this source streaming by homogeneous fluid flow. The field excited by the turbulent wake, string mixing zones, vortex formations is approximated by the system of trajectory distributed sourced on account of results of numerical calculation of spatial and energy characteristics of model two-dimensional problem describing the near field in exact setting. To simplest approximation it can believed that there's medium continuous mixing and internal gravity waves are excited with this mixed zone collapse; such field can be approximated by trajectory distributed vertically oriented mass dipole sources as fluid mixing resolves itself to mass part transfer from the mixed spot lower part to the upper part [6-10].

It is obvious, that the C problem solution is simpler than those of A and B.

Internal gravity waves propagation and excitation linear theory appears to be use efficient and with near field numerical calculation. Let us assume, that with $\xi = \xi_0 = 0$ (for definiteness) some medium perturbation localized near axis $\xi$ and not supposed to be small is set, so that this perturbation later evolution is to be calculated numerically. For that purpose it is necessary to select the region $|y| < M, |z| < N$, and at this rectangle borders it is necessary to lay down some boundary conditions, for example no-fluid-loss condition assuming $V = 0$ with $y = \pm M$ and $W = 0$ with $z = \pm N$. However when excited internal gravity waves reach with some $\xi = \xi_{max}$ designed rectangle borders, extraneous these borders-reflected waves origin absent in the real problem. Therefore numerical method solution can be calculated only with $\xi < \xi_{max}$. To damp reflected waves it is possible to lay down absorbing conditions at the borders, however such conditions are complex and are of approximate nature. The second solution continuation way is account of the circumstance waves approaching the border are relatively small in amplitude and are of linear nature. Therefore it is reasonable to separate field linear part. Let us consider the value $\xi_1 < \xi_{max}$ and let $(U_1, U_2, W_1), \eta_1$ are the velocity component and wave field elevation values with $\xi = \xi_1$. Let us assume these values as the initial data for linear equations and let us find corresponding solution $(U_1^*, U_2^*, W_1^*), \eta_1^*$. Let us later set up equations for the difference $\Delta U = (U_1^N - U_1^*, U_1^N - U_2^*, W_1^N - W_1^*), \Delta \eta = \eta_1^N - \eta_1^*$, where $(U_1^N, U_2^N, W_1^N), \eta_1^N$ is the solution of the initial nonlinear problem. This difference will be different from zero only in small neighborhood of the axis $\xi$, в in which calculated wave field is different form the linear one. Therefore it is possible to continue the solution $\Delta U, \Delta \eta$ with numerical methods to the region $\xi > \xi_{max}$. Whenever the function $\Delta U, \Delta \eta$ calculation region reaches the rectangle borders, it is possible again to distinguish the linear part and to continue calculations.

As internal gravity waves excitement, propagation in actual practice represents essentially nonlinear phenomenon, with some reasonable assumptions it is possible to linearize internal waves generation and propagation equations [6-10]. The present work studies linear approximations for internal gravity waves at large distances of an excitement source, in case when the source is substituted with a model. Therefore is interesting for example combination of model representations and semiempirical and possibly even experimental data. As internal waves equation solutions for a point excitement source were studied more closely in the first chapter, conceptually with knowledge of such solution it is possible to write the solution for the source vertically and horizontally smeared out, i.e. already having finite length and this again represent model reasoning.

The following approach is also possible. Let us assume, that there's an excitement source, which leaves turbulence wake (vortex formation region, mixing zone) generating internal waves in a certain manner [6-10]. Let us establish imaginary plane ($x = x_0$), parallel, assume, to excitement source motion trajectory and let us measure all wanted characteristics on this plane (Fig. 1). Then it is possible to solve the internal gravity waves propagation in linear approximation problem using plane data as initial ones. This way looks reasonable as good results of linear theory far from turbulence, mixing regions, i.e. from various physical nature excitement nonlocal sources should be expected. These nonlocal boundary data on this plane can be defined both experimentally and as a result of numerical calculations. In the same way in initial conditions much real information can be set, on the basis of which internal gravity waves far from nonlinearity region linear theory with $x > x_0$ should give satisfying results.

For the purpose of solving the problem of mathematic modeling of internal gravity waves generation by the region of partially mixed stratified medium it is possible to quote absorbing nonlocal boundary conditions, which take account of two essential from physical standpoint circumstances: first – linear theory is correct at large distances of perturbation sources; second – there are no other perturbation sources out mixing zone. Therefore use of these boundary conditions make it possible to describe diverging linear internal gravity waves excited by the region of mixed stratified medium (Fig..1).

**Nonlocal absorbing boundary conditions.**
Numerical algorithms for calculation of linear internal gravity waves far from excitement sources explicitly use or should use the results of hydrodynamics nonlinear problems exact numerical calculations as the formulation basis of for example physically correct boundary conditions [6-10]. However associated with this it is necessary to take account of the fact that with numerical solution of hydrodynamic nonlinear basic problems the topic of correct edge boundary conditions laying-down arises. One can assume, that at certain distances from excitement all sources it is admissible to linearize nonlinear problem, thus a possibility appears to use the results of internal gravity waves linear theory and with use of available exact analytical solutions to quote physically correct boundary conditions.

Let us for example consider the topic of numerical modeling of two-dimensional flow arising as a result of incompletely mixed fluid in stably vertically stratified medium region collapse (Fig.1). at large distances from the mixing region it is possible to use linear approximation and also the essential later on fact, that out of mixed fluid zone there are no excitement other sources. Use of these physically justified conditions imposes well-defined constraint on stream function and its derivatives at the exact numerical calculations region borders, which should be used for formulation of this problem numerical solution boundary conditions [6-10].

The numerical model of two-dimensional flow arising from the incompletely mixed fluid region collapse in vertically stably stratified medium is described by the Euler equation in Boussinesq approximation [6, 10]

$$\frac{\partial U}{\partial t} + U\frac{\partial U}{\partial x} + V\frac{\partial U}{\partial z} + \frac{1}{\rho_0(z)}\frac{\partial p}{\partial x} = 0$$

$$\frac{\partial V}{\partial t} + U\frac{\partial V}{\partial x} + V\frac{\partial V}{\partial z} + \frac{1}{\rho_0(z)}\frac{\partial p}{\partial z} + g\frac{\rho}{\rho_0(z)} = 0 \qquad (1)$$

$$\frac{\partial U}{\partial x} + \frac{\partial V}{\partial z} = 0$$

$$\frac{\partial \rho}{\partial t} + U\frac{\partial \rho}{\partial x} + V\frac{\partial \rho}{\partial z} + \frac{d\rho_0}{dz}V = 0$$

where $U, V$ are the velocity horizontal and vertical components accordingly, $\rho$ is density perturbation, $\rho_0(z)$ is quiescent fluid density [6,10]. By entering stream function $U = \frac{\partial \Psi}{\partial z}$, $V = -\frac{\partial \Psi}{\partial x}$ ($\Psi = const$ streamline) and vorticity $\omega = \frac{\partial U}{\partial z} - \frac{\partial V}{\partial x}$ this system is put in form

$$\frac{\partial \omega}{\partial t} + U\frac{\partial \omega}{\partial x} + V\frac{\partial \omega}{\partial z} = \frac{g}{\rho_0(z)}\frac{\partial \rho}{\partial x}$$

$$\frac{\partial^2 \Psi}{\partial x^2} + \frac{\partial^2 \Psi}{\partial z^2} = \omega$$

Linearized combined equations system is given by

$$\frac{\partial U}{\partial t} + \frac{1}{\rho_0(z)} \frac{\partial p}{\partial x} = 0$$

$$\frac{\partial V}{\partial t} + \frac{1}{\rho_0(z)} \frac{\partial p}{\partial z} + g \frac{\rho}{\rho_0(z)} = 0$$

$$\frac{\partial U}{\partial x} + \frac{\partial V}{\partial z} = 0$$

$$\frac{\partial \rho}{\partial t} + \frac{d \rho_0}{dz} V = 0$$

or

$$\frac{\partial \omega}{\partial t} = \frac{g}{\rho_0(z)} \frac{\partial \rho}{\partial x}$$

$$\frac{\partial^2 \Psi}{\partial x^2} + \frac{\partial^2 \Psi}{\partial z^2} = \omega \qquad (2)$$

$$\frac{\partial \rho}{\partial t} + \frac{d \rho_0}{dz} V = 0$$

With linearization and this system obtaining the second term of the equation

$$\frac{\partial \omega}{\partial t} = \frac{g}{\rho_0(z)} \frac{\partial \rho}{\partial x} + \frac{1}{\rho_0^2(z)} \frac{d \rho_0}{dz} \frac{\partial p}{\partial x}$$

was omitted. Let us evaluate the error of this approximation use, i.e. the ration between the second summand and the first one.

$$\frac{d\rho_0}{dz}\frac{\partial p}{\partial x}(g\rho_0(z)\frac{\partial \rho}{\partial x})^{-1} = -\frac{N^2(z)}{g^2}\frac{\partial p}{\partial x}(\frac{\partial \rho}{\partial x})^{-1} \qquad (2A)$$

Let

$$V = V_0 \exp(i(k_x x + k_z z - \phi t))$$

Then from the system (1) third equation we have

$$U = -V_0 k_z \exp(i(k_x x + k_z z - \phi t))/k_x$$

Later it is possible to obtain the following relations

$$\omega = \frac{\partial U}{\partial z} - \frac{\partial V}{\partial x} = -i(k_x^2 + k_z^2)\exp(i(k_x x + k_z z - \phi t))/k_x$$

$$\frac{\partial \rho}{\partial x} = \frac{\partial \omega}{\partial t}\rho_0(z)/g = -\phi(k_x^2 + k_z^2)\rho_0(z)\exp(i(k_x x + k_z z - \phi t))/g k_x$$

$$\frac{\partial p}{\partial x} = -i\phi k_z \rho_0(z)\exp(i(k_x x + k_z z - \phi t))/k_x$$

As a result the formula (2A) can be given as

$$\frac{N^2(z)}{g^2}\frac{\partial p}{\partial x}(\frac{\partial \rho}{\partial x})^{-1} = N^2(z)k_z/g(k_x^2 + k_z^2)$$

It is obvious that for real oceanic scales ($N(z) \approx 10^{-2} c^{-1}, g \approx 10 \, м/c^{-2}$) this expression is much less than unity [7-9].

Let us show, that the stream function $\Psi$, defined by this linearized combined equations satisfies the internal gravity linear waves equation. Indeed let us differentiate the system (2) first equation with respect to time

$$\frac{\partial^2 \omega}{\partial t^2} = \frac{g}{\rho_0}\frac{\partial^2 \rho}{\partial x \partial t} = -\frac{g}{\rho_0(z)}\frac{d\rho_0}{dz}\frac{\partial V}{\partial x} = N^2(z)\frac{\partial V}{\partial x} = -N^2(z)\frac{\partial^2 \Psi}{\partial x^2}$$

where $N^2(z) = -\dfrac{g}{\rho_0(z)}\dfrac{d\rho_0}{dz}$ is the Brunt-Vaisala frequency, which is the main parameter defining internal gravity waves characteristics [7-9].

As

$$\frac{\partial^2 \omega}{\partial t^2} = \frac{\partial^2}{\partial t^2}\left(\frac{\partial^2 \Psi}{\partial x^2} + \frac{\partial^2 \Psi}{\partial z^2}\right)$$

we have

$$\frac{\partial^2}{\partial t^2}\left(\frac{\partial^2 \Psi}{\partial x^2} + \frac{\partial^2 \Psi}{\partial z^2}\right) + N^2(z)\frac{\partial^2 \Psi}{\partial x^2} = 0 \tag{3}$$

Let us consider the function $\Omega = \dfrac{\omega}{N^2(z)}$, i.e. $\omega = \Omega N^2(z)$

$$\frac{\partial \Omega}{\partial t} N^2(z) = \frac{g}{\rho_0(z)}\frac{\partial \rho}{\partial x}$$

$$\frac{\partial^2 \Psi}{\partial x^2} + \frac{\partial^2 \Psi}{\partial z^2} = \Omega N^2(z) \tag{4}$$

Let us differentiate the equation (4) with respect to time

$$\frac{\partial^2 \Omega}{\partial t^2} N^2(z) = \frac{g}{\rho_0}\frac{\partial^2 \rho}{\partial x \partial t} = -\frac{g}{\rho_0(z)}\frac{d\rho_0}{dz}\frac{\partial V}{\partial x} = N^2(z)\frac{\partial V}{\partial x} = -N^2(z)\frac{\partial^2 \Psi}{\partial x^2}$$

Later we have

$$\frac{\partial^2 \Omega}{\partial t^2} = -\frac{\partial^2 \Psi}{\partial x^2}$$

$$\frac{\partial^4 \Omega}{\partial t^2 \partial x^2} = -\frac{\partial^4 \Psi}{\partial x^4}$$

$$\frac{\partial^4 \Omega}{\partial t^2 \partial z^2} = -\frac{\partial^4 \Psi}{\partial z^2 \partial x^2}$$

Let us combine two equations and as a result we obtain

$$\frac{\partial^2}{\partial t^2}(\frac{\partial^2 \Omega}{\partial x^2} + \frac{\partial^2 \Omega}{\partial z^2}) = -\frac{\partial^2}{\partial x^2}(\frac{\partial^2 \Psi}{\partial x^2} + \frac{\partial^2 \Psi}{\partial z^2}) = -\frac{\partial^2}{\partial x^2}(\Omega N^2(z))$$

or

$$\frac{\partial^2}{\partial t^2}(\frac{\partial^2 \Omega}{\partial x^2} + \frac{\partial^2 \Omega}{\partial z^2}) + N^2(z)\frac{\partial^2 \Omega}{\partial x^2} = 0$$

i.e. streamline function $\Psi$ and the function $\Omega = \dfrac{\omega}{N^2(z)}$ (in case $N^2(z) = const$ and vorticity $\omega$) itself satisfy the internal gravity waves main equation [1-5, 8]. As a rule the condition $\Psi = 0$ is used as boundary conditions. However as it will be shown below, if the streamline function satisfies internal gravity waves equation, there is well-defined dependence between the functions $\Psi$ and $\dfrac{\partial \Psi}{\partial x}$ which should be used as corresponding boundary conditions for correct numerical solution of the stratified mediums dynamics problems.

**Formulation of nonlocal boundary condition for finite stratified medium.**

Let us consider the case $N(z) = const$. The Green's function $G$ for internal gravity waves in stratified medium $0 < z < H$ finite layer is defined from the equation [1-5, 8]

$$\frac{\partial^2}{\partial t^2}\left(\frac{\partial^2}{\partial x^2} + \frac{\partial^2}{\partial z^2}\right)G + N^2(z)\frac{\partial^2}{\partial x^2}G = \delta(t)\delta(x)\delta(z - z')$$

where $z'$ is perturbation source depth, $\delta(x)$ is delta function of Dirac. In dimensionless coordinates $t^* = Nt$, $x^* = \dfrac{x}{H}$, $z^* = \dfrac{z}{H}$, $G^* = GN$ we have

$$\frac{\partial^2}{\partial t^{*2}}\left(\frac{\partial^2}{\partial x^{*2}} + \frac{\partial^2}{\partial z^{*2}}\right)G^* + \frac{\partial^2}{\partial x^{*2}}G^* = \delta(t^*)\delta(x^*)\delta(z^* - z'^*)$$

Later we omit the index $*$

$$\frac{\partial^2}{\partial t^2}\Delta G + \frac{\partial^2}{\partial x^2}G = \delta(t)\delta(x)\delta(z - z')$$

$$\Delta = \frac{\partial^2}{\partial x^2} + \frac{\partial^2}{\partial z^2}$$

Boundary and initial conditions for the function $G$, defininition are of the form

$$G(z,x,t) = \frac{\partial G(z,x,t)}{\partial t} \equiv 0 \text{ при } t < 0; \; G\big|_{z=0} = G\big|_{z=1} = 0$$

$$G < \infty \quad \text{при } \sqrt{x^2 + y^2} \to \infty$$

The Green's function $G$, satisfying these conditions, is constructed in [1-5]

$$G(z,x,t) = \frac{\theta(t)}{4\pi}\int_0^{\pi/2} \ln\left|\frac{\cos(\pi x \, tg\varphi) - \cos\pi(z-z')}{\cos(\pi x \, tg\varphi) - \cos\pi(z+z')}\right| \cdot \frac{\sin(t \sin\varphi)}{\sin\varphi} d\varphi$$

Let us substitute $tg\varphi = u$

$$G = \frac{\theta(t)}{4\pi}\int_0^\infty \ln\left|\frac{\cos(\pi xu) - \cos\pi(z-z')}{\cos(\pi xu) - \cos\pi(z+z')}\right| \cdot \frac{\sin\dfrac{tu}{\sqrt{1+u^2}}}{u\sqrt{1+u^2}} du \tag{5}$$

The Green's function $G$ possesses the following properties: $G(z,x,t) \to 0$ with $t \to +0$

$$\frac{\partial G}{\partial t} = \frac{1}{8} \ln \frac{ch\pi x - \cos\pi(z-z')}{ch\pi x - \cos\pi(z+z')} \quad \text{with } t \to +0 \tag{6}$$

To prove the formula (6) let us differentiate t function $G(z,x,t)$ and let us substitute $t=0$, then we obtain

$$\left.\frac{\partial G}{\partial t}\right|_{t=0} = \frac{\theta(t)}{4\pi} \operatorname{Re} \int_0^\infty \ln\left|\frac{\cos(\pi xu) - \cos\pi(z-z')}{\cos(\pi xu) - \cos\pi(z+z')}\right| \cdot \frac{du}{1+u^2} \tag{7}$$

Let us later consider the integral (7) on complex plane $u$

$$\left.\frac{\partial G}{\partial t}\right|_{t=0} = \frac{\theta(t)}{8\pi} \operatorname{Re} \int_{-\infty+i\varepsilon}^{\infty+i\varepsilon} \left[\ln \frac{\cos(\pi xu) - \cos\pi(z-z')}{\cos(\pi xu) - \cos\pi(z+z')}\right] \cdot \frac{du}{1+u^2}$$

By closing contour in the upper half-plane and with residue in the simple pole $u=i$ it is possible to obtain the formula (6).

With $x \to 0$ и $(z-z') \to 0$, the Green's function time derivative is logarithmically infinite

$$\left.\frac{\partial G(z,x,t)}{\partial t}\right|_{\substack{t\to+0\\x\to 0\\(z-z')\to 0}} = \frac{1}{8}\ln(x^2 + (z-z')^2)$$

Let us write the Green's formula for internal gravity waves equation

$$\int_0^H dz \int_{-\infty}^\infty dt \int_0^\infty dx\, G\left(\frac{\partial^2}{\partial t^2}\Delta\Psi + \frac{\partial^2 \Psi}{\partial x^2}\right) =$$

$$= \int_0^H dz \int_{-\infty}^\infty dt \int_0^\infty dx\, \Psi\left(\frac{\partial^2}{\partial t^2}\Delta G + \frac{\partial^2 G}{\partial x^2}\right) +$$

$$+\int_0^H dz \int_{-\infty}^{\infty} \left[\left(\frac{\partial^2}{\partial t^2}+1\right)G\cdot\frac{\partial \Psi}{\partial x} - \Psi\left(\frac{\partial^2}{\partial t^2}+1\right)\frac{\partial G}{\partial x}\right]_{x=0} \quad (8)$$

We assume, that in the region $x > x_0$ later without loss of generality $x_0 \equiv 0$) there are no sources, then it is obvious, that the Green's function $G$ satisfies the equation

$$\frac{\partial^2}{\partial t^2}\Delta G + \frac{\partial^2 G}{\partial x^2} = 0$$

As the streamline function $\Psi$ satisfies the internal gravity waves equation (3), then with $x = 0$ there is single-valued connection between the streamline function $\Psi$ and its derivative $\frac{\partial \Psi}{\partial x}$, and it is no longer possible to prescribe arbitrarily these functions values. Therefore with numerical solution of the problem (2) it is necessary to use the following integral relation as boundary conditions

$$\int_0^H dz \int_{-\infty}^{\infty} (F_1 \cdot \frac{\partial \Psi}{\partial x} - \Psi F_2) dt = 0 \quad (9)$$

$$F_1(z,t) = \left(\frac{\partial^2}{\partial t^2}+1\right)G\bigg|_{x=0}$$

$$F_2(z,t) = \left(\frac{\partial^2}{\partial t^2}+1\right)\frac{\partial G}{\partial x}\bigg|_{x=0}$$

The function $F_1(z,t)$ is as follows

$$F_1(z,t) = \frac{1}{4\pi}\ln\frac{1-\cos\pi(z-z')}{1-\cos\pi(z+z')}\cdot\left\{\frac{\pi}{2}\delta(t) - \theta(t)\int_0^1 \frac{u\sin tu}{\sqrt{1-u^2}}du + \theta(t)\int_0^1 \frac{\sin tu}{u\sqrt{1-u^2}}du\right\} \quad (10)$$

The function $F_1(z,t)$, after operator application $\left(\dfrac{\partial^2}{\partial t^2}+1\right)$ to the function $G\big|_{x=0}$ and substitution $\sin\varphi = u$ can be also given as

$$G\big|_{x=0} = \frac{\theta(t)}{4\pi}\ln\frac{1-\cos\pi(z-z')}{1-\cos\pi(z+z')}\cdot\frac{\pi}{2}\int_0^t J_0(\tau)d\tau$$

$$\left(\frac{\partial^2}{\partial t^2}+1\right)G\big|_{x=0} =$$

$$=\frac{1}{8}\ln\frac{1-\cos\pi(z-z')}{1-\cos\pi(z+z')}\cdot\left\{\frac{d^2\theta(t)}{dt^2}\int_0^t J_0(\tau)d\tau + 2\frac{d\theta(t)}{dt}J_0(t) - \theta(t)J_1(t) + \theta(t)\cdot\int_0^t J_0(\tau)d\tau\right\}=$$

$$=\frac{1}{8}\ln\frac{1-\cos\pi(z-z')}{1-\cos\pi(z+z')}\cdot\left\{\frac{d\delta(t)}{dt}\int_0^t J_0(\tau)d\tau + 2\delta(t)J_0(t) - \theta(t)J_1(t) + \theta(t)\cdot\int_0^t J_0(\tau)d\tau\right\}=$$

$$=\frac{1}{8}\ln\frac{1-\cos\pi(z-z')}{1-\cos\pi(z+z')}\cdot\left\{\delta(t)J_0(t) - \theta(t)J_1(t) + \theta(t)\cdot\int_0^t J_0(\tau)d\tau\right\}$$

where $J_0(\tau)$, $J_1(t)$ is Bessel's function [11-13]. Thus the expression (10) can be given as

$$F_1(z,t) = \frac{1}{8}\ln\frac{1-\cos(\pi(z-z'))}{1-\cos(\pi(z+z'))}\cdot\left\{\delta(t) - \theta(t)J_1(t) + \theta(t)\cdot\int_0^t J_0(\tau)d\tau\right\}$$

Let us find the expression for the function $F_2(z,t)$

$$F_2(z,t) = I_1 + I_2 \tag{11}$$

$$I_1 = \frac{\delta(t)}{8\pi} \int_{-\infty}^{\infty} \frac{u \sin(\pi x u)}{1+u^2} \left[ \frac{\cos(\pi(z+z')) - \cos \pi(z-z')}{(\cos(\pi x u) - \cos \pi(z+z'))(\cos(\pi x u) - \cos \pi(z-z'))} \right] du$$

$$I_2 = \frac{\theta(t)}{8\pi} \int_{-\infty}^{\infty} \frac{\sin(\pi x u) \sin \frac{tu}{\sqrt{1+u^2}}}{(1+u^2)^{3/2}} \left[ \frac{\cos \pi(z+z') - \cos \pi(z-z')}{(\cos(\pi x u) - \cos \pi(z+z'))(\cos(\pi x u) - \cos \pi(z-z'))} \right] du$$

The integrals (11) are understood in terms of the principal value around the poles on complex plane real axis $u$. Then for calculation of the expression $I_1$ we close the contour in the upper half-plane and using the Jordan lemma [11] with residue in the simple pole $u = i$, it is possible to obtain

$$I_1 = \frac{\delta(t)}{8} \cdot -sh\pi x \left[ \frac{1}{ch\pi x - \cos \pi(z-z')} - \frac{1}{\cos \pi x - \cos \pi(z-z')} \right]$$

With $x \to 0$ and $z - z' \to 0$ we have

$$I_1 = \frac{\delta(t)}{4\pi} \frac{x}{x^2 + (z-z')^2} \to \frac{\delta(t)\delta(z-z')}{4}$$

The summand $I_2 : I_2 \to 0$ should now be studied with $x \to 0$, $z \neq z_1$. Let us consider behavior $I_2$ with $x \to 0$, $z - z' \to 0$ and small t. At the first approximation the summand $I_2$ can be given as

$$I_2 \approx \frac{\theta(t) \cdot tx}{8} \int_{-\infty}^{\infty} \frac{-1}{\cos(\pi x u) - \cos \pi(z-z')} \cdot \frac{u^2 du}{(1+u^2)^2} \tag{12}$$

The integral (12) is still understood in terms of the principle value. On the complex plane u let us close the contour in the upper half-plane and let us make residue at the point $u = i$, which is the pole, but already of quadric surface. As a result we obtain

$$I_2 \approx -\frac{\theta(t)t}{8\pi}\left(\frac{x}{x^2+(z-z')^2} - \frac{2x^3}{\left(x^2+(z-z')^2\right)^2}\right)$$

As

$$\int_{-\infty}^{\infty}\frac{xdx}{x^2+(z-z')^2} = \pi$$

$$\int_{-\infty}^{\infty}\frac{x^3}{\left(x^2+(z-z')^2\right)^2} = \frac{\pi}{2}$$

at the first approximation it is possible to obtain

$$I_2 \approx -\frac{\theta(t)t}{8}(\delta(z-z') - \delta(z-z')) \approx 0$$

It is possible to show, that the second and other higher approximations of the summand $I_2$ are equal to zero, however we do not cite the proof because of its awkwardness.

As a result we have

$$F_1(z,t) = \frac{1}{8}\ln\frac{1-\cos\pi(z-z')}{1-\cos\pi(z+z')} \cdot \left\{\delta(t) - \theta(t)J_1(t) + \theta(t)\cdot\int_0^t J_0(\tau)d\tau\right\}$$

$$F_2(z,t) = \frac{\delta(t)\delta(z-z')}{4} + 0$$

**Formulation of nonlocal boundary condition for stratified infinite depth medium.**

Plane $x=0$ bounded half-space is considered with Brunt-Vaisala frequency constant distribution $N(z) = const$. In the half-space $x \geq 0$ the streamline function $\Psi$ satisfies linear internal gravity waves equation [1-5, 8]

$$\frac{\partial^2}{\partial t^2}\Delta\Psi + N^2(z)\frac{\partial^2\Psi}{\partial x^2} = 0$$

In the half-space with $N(z) = const$ it is possible to write definite representation for the Green's function $\Gamma$ satisfying the equation

$$\frac{\partial^2}{\partial t^2}\Delta\Gamma + N^2(z)\frac{\partial^2\Gamma}{\partial x^2} = \delta(x)\delta(z-z_0)\delta(t)$$

which takes the form [1-5]

$$\Gamma = \theta(t)\int_0^N \frac{\sin\omega t \ln\left|(N^2-\omega^2)(z-z_0)^2 - \omega^2 x^2\right|}{\omega\sqrt{N^2-\omega^2}}d\omega$$

Let us later write the Green's formula for internal gravity waves equation and as a result we obtain

$$\int_{-\infty}^{\infty}dt\int_0^{\infty}dx\int_{-\infty}^{\infty}dz\,\Gamma\left(\frac{\partial^2}{\partial t^2}\Delta\Psi + N^2\frac{\partial^2\Psi}{\partial x^2}\right) = \int_{-\infty}^{\infty}dt\int_0^{\infty}dx\int_{-\infty}^{\infty}dz\,\Psi\left(\frac{\partial^2}{\partial t^2}\Delta\Gamma + N^2\frac{\partial^2\Gamma}{\partial x^2}\right) -$$

$$-\int_{-\infty}^{\infty}dt\int_{-\infty}^{\infty}dz\left\{\left(N^2 + \frac{\partial^2}{\partial t^2}\right)\Gamma\frac{\partial\Psi}{\partial x} - \Psi\left(\frac{\partial^2}{\partial t^2} + N^2\right)\frac{\partial\Gamma}{\partial x}\right\}\bigg|_{x=0} = 0$$

In this Green's formula the limits of integration with respect ot z, t, x are infinite, herewith we consider, that at infinity all functions go to zero. With $t \to -\infty$ one integrand function goes to zero, with $t \to +\infty$ - another, but no two simultaneously. Then with non-zero integrating only plane $x = 0$ contribution remains. As the functions $\Psi$, $\Gamma$ satisfy internal gravity waves equation

$$\int_{-\infty}^{\infty}dt\int_{-\infty}^{\infty}dz\left\{\left(N^2 + \frac{\partial^2}{\partial t^2}\right)\Gamma\frac{\partial\Psi}{\partial x} - \Psi\left(\frac{\partial^2}{\partial t^2} + N^2\right)\frac{\partial\Gamma}{\partial x}\right\}\bigg|_{x=0} = 0$$

or

$$\int_{-\infty}^{\infty}dz\int_{-\infty}^{\infty}(F_1\cdot\frac{\partial\Psi}{\partial x} - \Psi F_2)dt = 0 \qquad (13)$$

$$F_1(z,t) = \left(\frac{\partial^2}{\partial t^2} + N^2\right)\Gamma\bigg|_{x=0}$$

$$F_2(z,t) = \left(\frac{\partial^2}{\partial t^2} + N^2\right)\frac{\partial\Gamma}{\partial x}\bigg|_{x=0}$$

Let us calculate these two functions

$$F_1(z,t) = \left(\frac{\partial^2}{\partial t^2} + N^2\right)\Gamma\bigg|_{x=0}$$

We have

$$\left(\frac{\partial^2}{\partial t^2} + N^2\right)\Gamma = \delta(t)\int_0^N \frac{\ln\left|(N^2-\omega^2)(z-z_0)^2 - \omega^2 x^2\right|}{\sqrt{N^2-\omega^2}} d\omega$$

$$+ \theta(t)\int_0^N \sin\omega t \frac{\sqrt{N^2-\omega^2}}{\omega} \ln\left|(N^2-\omega^2)(z-z_0)^2 - \omega^2 x^2\right| d\omega \equiv I_1 + I_2$$

Let us consider the summand $I_1$

$$I_1\big|_{x=0} = \delta(t)\int_0^N \frac{\ln\left|(N^2-\omega^2)(z-z_0)^2\right|}{\sqrt{N^2-\omega^2}} d\omega$$

Using substitution: $\omega = \dfrac{\kappa N}{\sqrt{\kappa^2+1}}$, we obtain

$$N^2 - \omega^2 = \frac{N^2}{\kappa^2+1}, \quad d\omega = \frac{N}{(\kappa^2+1)^{3/2}} d\kappa$$

$$I_1\big|_{x=0} = \delta(t)\int_0^\infty \frac{\ln(N^2(z-z_0)^2) - \ln(\kappa^2+1)}{\kappa^2+1} d\kappa =$$

$$\int_0^\infty \frac{\ln(N^2(z-z_0)^2)}{\kappa^2+1} d\kappa = \ln(N^2(z-z_0)^2)\cdot\frac{\pi}{2}$$

$$\int_0^\infty \frac{\ln(\kappa^2+1)}{\kappa^2+1} d\kappa = \frac{1}{2}\int_{-\infty}^\infty \frac{\ln(\kappa+i)}{\kappa^2+1} + \frac{\ln(\kappa-i)}{\kappa^2+1} d\kappa = \pi\ln 2$$

So

$$I_1\big|_{x=0} = \delta(t)\pi \ln\frac{N|z-z_0|}{2}$$

Let us now consider the summand $I_2$

$$I_2\big|_{x=0} = \theta(t)\int_0^N \sin\omega t \frac{\sqrt{N^2-\omega^2}}{\omega}\ln(z-z_0)^2 d\omega +$$

$$+ \theta(t)\int_0^N \sin\omega t \frac{\sqrt{N^2-\omega^2}}{\omega}\ln|N^2-\omega^2| d\omega \equiv B_1 + B_2$$

Let us study the integral $B_1$, having made substitution $\omega = N\sin\varphi$

$$B_1 = \theta(t)N\int_0^{\pi/2}\sin(Nt\sin\varphi)\frac{\cos^2\varphi}{\sin^2\varphi}\ln(z-z_0)^2 d\varphi =$$

$$= \theta(t)N\ln(z-z_0)^2\left\{\int_0^{\pi/2}\frac{\sin(Nt\sin\varphi)}{\sin\varphi}d\varphi - \int_0^{\pi/2}\sin(Nt\sin\varphi)\sin\varphi d\varphi\right\} =$$

$$= \theta(t)N\ln(z-z_0)^2\left\{\frac{\pi N}{2}\int_0^t J_0(N\tau)d\tau - \frac{\pi}{2}J_1(Nt)\right\} =$$

$$= \theta(t)\pi N\ln|z-z_0|\left\{N\int_0^t J_0(N\tau)d\tau - J_1(Nt)\right\}$$

The integral $B_2$ (after substitution $\omega = N\sin\varphi$) is

$$B_2 = \theta(t)N\int_0^{\pi/2}\sin(Nt\sin\varphi)\frac{\cos^2\varphi}{\sin^2\varphi}\ln|N^2\cos^2\varphi|d\varphi$$

Later it is possible to obtain

$$I_2 = B_1 + B_2$$

$$B_1 = \theta(t)2\pi N\ln N|z-z_0|\left\{N\int_0^t J_0(N\tau)d\tau - J_1(Nt)\right\}$$

$$B_2 = \theta(t)N\int_0^{\pi/2}\sin(Nt\sin\varphi)\frac{\cos^2\varphi}{\sin\varphi}\ln\cos^2\varphi d\varphi =$$

$$= \theta(t)N\int_0^{\pi/2}\sin(Nt\sin\varphi)\frac{\ln\cos^2\varphi}{\sin\varphi}d\varphi - \theta(t)N\int_0^{\pi/2}\sin(Nt\sin\varphi)\sin\varphi\ln(\cos^2\varphi)d\varphi \equiv A_1 + A_2$$

To express the summands $A_1$ и $A_2$, let us first consider the integral

$$I(\tau) = \theta(t)N \int_0^{\pi/2} \cos(Nt \sin\varphi) \ln(\cos^2 \varphi) d\varphi = \theta(t)N \int_0^{\pi/2} \cos(\tau \sin\varphi) \ln(\cos^2 \varphi) d\varphi$$

To calculate it let us consider the following representation of the Bessel's function (Poisson integral representation) [11-13]

$$J_\nu(\tau) = \frac{2\left(\frac{x}{2}\right)^\nu}{\sqrt{\pi}\,\Gamma(\nu+\frac{1}{2})} \int_0^{\pi/2} \cos(\tau \sin\varphi) \cos^{2\nu}\varphi \, d\varphi$$

where $\Gamma(z)$ is gamma function [11-13], or

$$\frac{\sqrt{\pi}}{2} \frac{J_\nu(\tau)\Gamma(\nu+\frac{1}{2})}{\left(\frac{x}{2}\right)^\nu} = \int_0^{\pi/2} \cos(\tau \sin\varphi) \cos^{2\nu}\varphi \, d\varphi$$

Let us differentiate both members of the last-mentioned equation in respect to $\nu$ and let us assume $\nu = 0$. Then we obtain

$$\int_0^{\pi/2} \cos(\tau \sin\varphi) \ln(\cos^2 \varphi) d\varphi = \frac{\sqrt{\pi}}{2} \frac{d}{d\nu} \left( \frac{J_\nu(\tau)\Gamma(\nu+\frac{1}{2})}{\left(\frac{x}{2}\right)^\nu} \right)_{\nu=0}$$

To calculate the right-hand member let us use the following relations [11-13]

$$\left.\frac{dJ_\nu(\tau)}{d\nu}\right|_{\nu=0} = \frac{\pi}{2} Y_0(x)$$

$$\Gamma\left(\frac{1}{2}\right) = \sqrt{\pi}, \quad \Gamma'\left(\frac{1}{2}\right) = \sqrt{\pi}(-C - \ln 4)$$

where $Y_0(x)$ is a zeroth-order Neumann function, $C \approx 0.577$ is the Euler's constant [11-13].
Finally we obtain

$$I(\tau) = \theta(t) N \frac{\pi}{4} (\pi Y_0(Nt) - J_0(Nt)(2C + \ln 4tN^2))$$

$$A_1(\tau) = \int_0^\tau I(u)du$$

$$A_2(\tau) = -\frac{dI(\tau)}{d\tau} = -\theta(t)N\frac{\pi}{4}\left(\frac{2J_0(Nt)}{Nt} + \pi Y_1(Nt) - J_1(Nt)(2C + \ln 4t^2 N^2)\right)$$

As a result we have

$$F_1(z,t) = \left(\frac{\partial^2}{\partial t^2} + N^2\right)\Gamma\bigg|_{x=0} = \delta(t)\pi \ln\frac{N(z-z_0)}{2} + \theta(t)N\left\{2\pi \ln(N|z-z_0|)\left[\int_0^\tau J_0(u)du - J_1(\tau)\right] + \right.$$

$$\left. + \frac{\pi}{2}\int_0^\tau \left(\frac{\pi}{2}Y_0(u) - J_0(u)(C + \ln(2u))\right)du + \frac{\pi}{2}\left(\frac{J_0(\tau)}{\tau} + \frac{\pi}{2}Y_1(\tau) - J_1(\tau)(C + \ln(2\tau))\right)\right\}$$

where $\tau = Nt$, C is the Euler's constant, $J_0(u)$, $J_1(u)$ are the Bessel's zeroth- and first-order functions, $Y_0(u)$, $J_1(u)$ are the Neumann zeroth- and first-order functions accordingly [11-13].

Let us show later, that the function $F_2(z,t) = \left(\frac{\partial^2}{\partial t^2} + N^2\right)\frac{\partial \Gamma}{\partial x}\bigg|_{x=0}$ is

$$\left(\frac{\partial^2}{\partial t^2} + N^2\right)\frac{\partial \Gamma}{\partial x}\bigg|_{x\to 0} = \pi^2 \delta(t)\delta(z-z_0)$$

Indeed

$$\left(\frac{\partial^2}{\partial t^2} + N^2\right)\frac{\partial \Gamma}{\partial x}\bigg|_{x\to 0} = \delta(t)\int_0^N \frac{-2\omega^2(x-x_0)}{\sqrt{N^2-\omega^2}((N^2-\omega^2)(z-z_0)^2 - \omega^2(x-x_0)^2)}d\omega +$$

$$+ \theta(t)\int_0^N \sin\omega t \frac{\sqrt{N^2-\omega^2}}{\omega} \frac{-2\omega^2(x-x_0)}{((N^2-\omega^2)(z-z_0)^2 - \omega^2(x-x_0)^2)}d\omega \equiv \delta(t)I_1 + \theta(t)I_2$$

Let us consider the summand $I_1$. Let us substitute: $\omega = \frac{\kappa N}{\sqrt{\kappa^2+1}}$, as a result we obtain

$$I_1 = -\int_{-\infty}^{\infty} \frac{\kappa^2(x-x_0)}{(\kappa^2+1)((z-z_0)^2 - \kappa^2(x-x_0)^2)}d\kappa$$

where the integral is understood in term of the principle value. By closing the contour in the upper half-plane and defining residue at the point $k = i$, we obtain

$$I_1 = -2\pi i \frac{-(x-x_0)}{2i((z-z_0)^2 + (x-x_0)^2)} = \pi \frac{(x-x_0)}{(z-z_0)^2 + (x-x_0)^2} = \pi^2 \delta(z-z_0)$$

The proof of $I_2 \to 0$ with $x \to x_0$ is similar to above considered case of stratified medium finite layer. Indeed the integral $I_2$ with small t reduces to the expression

$$I_2 \approx -N^2 t \int_{-\infty}^{\infty} \frac{\kappa^2 (x-x_0)}{(\kappa^2 + 1)((z-z_0)^2 - \kappa^2 (x-x_0)^2)} d\kappa$$

By closing the contour on top and defining residue at the point $k = i$, which is the second-order pole and at the first approximation we obtain $I_2 \approx 0$. it is possible to show, that the second and other higher approximation of the summand $I_2$ are also equal to zero, but we do not cite the proof because of its awkwardness.

**Discussion.**

Thus for purpose of the numerical solution of the mathematic modeling of internal gravity waves generation by the region of partially mixed stratified medium it is possible to quote absorbing nonlocal boundary conditions in form of (9), (13), which take account of two essential from physical standpoint circumstances: the first – at large distances of perturbation sources linear theory is correct; the second – out of mixing zone there are no other perturbation sources. Therefore use of these boundary conditions makes it possible to describe diverging linear internal gravity waves excited by the region of mixed stratified medium. Using obtained results it is possible to obtain the answer to the question of principle of how to define later dynamics of internal gravity waves nonharmonic trains far from these perturbation sources in respect of some set (numerically, experimentally defined) stratified medium parameters distribution, assuming adequacy of wave dynamics linear model use at large distances.

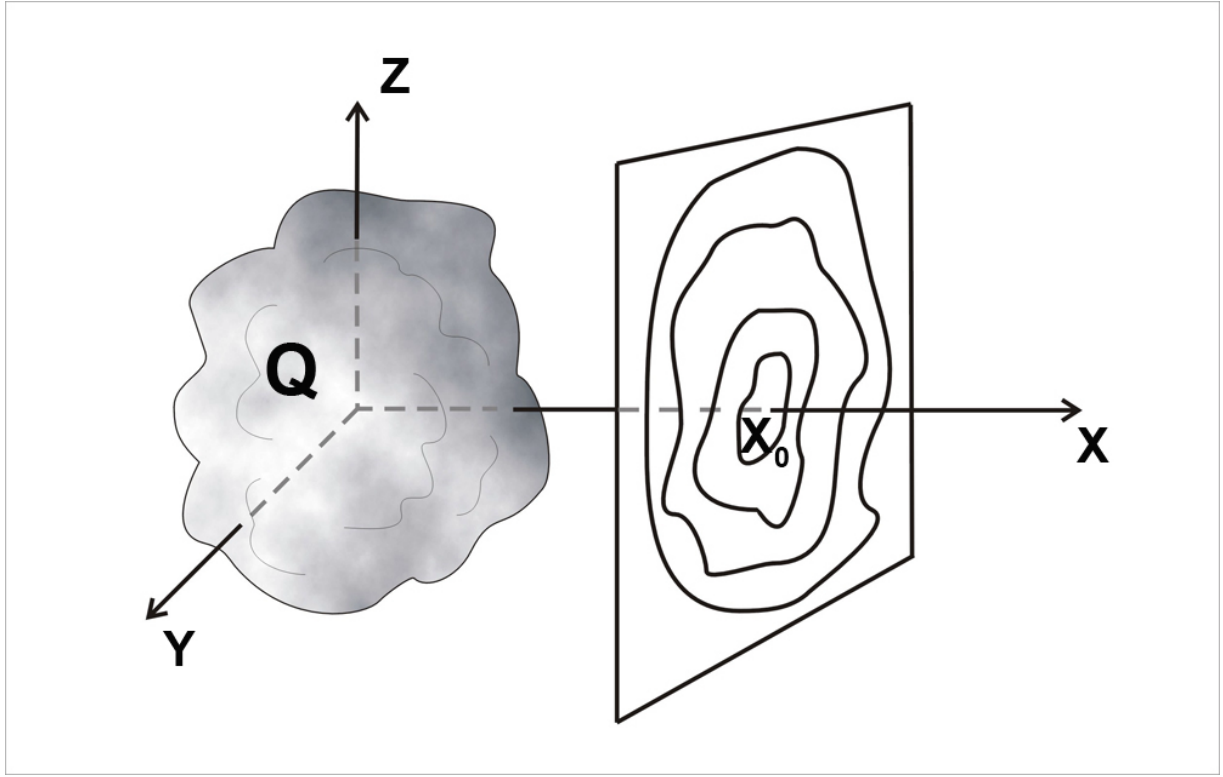

Fig.1. Nonlocal boundary conditions for mathematic modeling of stratified medium wave dynamics far from perturbation sources. Spatial region $Q$ - numerical solution of complete nonlinear problem, $x > x_0$ - linear theory applicability limit, the plane $x = x_0$ - boundary conditions laying-down for correct solution of wave dynamics problem.

**References.**

1. Bulatov, V.V. and Vladimirov, Y.V. 2007 Internal gravity waves: theory and applications. Moscow, Nauka, 304 pp.

2. Bulatov, V.V. and Vladimirov, Y.V. 2006 General problems of the internal gravity waves linear theory // Cornell University Library, 2006, E-Print Archive, Paper ID: physics/0609236, http://arxiv.org/abs/physics/0609236